\documentclass[conference]{IEEEtran}

\usepackage{booktabs} 

\usepackage{graphicx}
\usepackage{balance}  
\usepackage{times}
\usepackage{amsmath,amsthm,epsfig}

\usepackage[ruled,vlined,linesnumbered,commentsnumbered]{algorithm2e}
\usepackage{color}
\usepackage{booktabs}
\usepackage{multicol}
\usepackage{epsfig}
\usepackage{epstopdf}
\usepackage{url}
\usepackage{float}
\usepackage{subcaption}
\usepackage[labelfont=bf,skip=7pt]{caption}
\usepackage{changepage}
\usepackage{soul}
\usepackage{wrapfig}
\usepackage{ragged2e}
\usepackage{todonotes}
\usepackage{multirow}

\usepackage{sistyle}
\SIthousandsep{,}

\newtheorem{lemma}{Lemma}
\newtheorem{corollary}{Corollary}

\newcommand{\comment}[1]{{\color{red}#1}}
\newcommand{\modified}[1]{{\color{blue}#1}}
\newcommand{\remove}[1]{}

\newcommand{\cliques}{\mathcal{C}}
\newcommand{\tomita}{{\tt TTT}\xspace}
\newcommand{\partomita}{{\tt ParTTT}\xspace}
\newcommand{\parmce}{{\tt ParMCE}\xspace}
\newcommand{\parmcedegree}{{\tt ParMCEDegree}\xspace}
\newcommand{\parmcedegen}{{\tt ParMCEDegen}\xspace}
\newcommand{\parmcetriangle}{{\tt ParMCETri}\xspace}
\newcommand{\parpivot}{{\tt ParPivot}\xspace}
\newcommand{\cand}{{\tt cand}\xspace}
\newcommand{\fini}{{\tt fini}\xspace}
\newcommand{\pivot}{{\tt pivot}\xspace}
\newcommand{\ext}{{\tt ext}\xspace}
\newcommand{\rank}{{\tt rank}\xspace}
\newcommand{\intersect}{{\tt intersect}\xspace}

\newcommand{\greedybb}{{\tt GreedyBB}\xspace}
\newcommand{\hashing}{{\tt Hashing}\xspace}
\newcommand{\bkd}{{\tt BKDegeneracy}\xspace}
\newcommand{\cliqueenum}{{\tt Clique Enumerator}\xspace}
\newcommand{\peamc}{{\tt Peamc}\xspace}
\newcommand{\peco}{{\tt PECO}\xspace}
\newcommand{\pecodegree}{{\tt PECODegree}\xspace}
\newcommand{\pecodegen}{{\tt PECODegen}\xspace}
\newcommand{\pecotri}{{\tt PECOTri}\xspace}
\newcommand{\pecoshared}{{\tt PECOS}\xspace}
\newcommand{\pecoshareddegree}{{\tt PECOS-Degree}\xspace}
\newcommand{\pecoshareddegen}{{\tt PECOS-Degen}\xspace}
\newcommand{\pecosharedtri}{{\tt PECOS-Tri}\xspace}

\begin{document}
\title{Shared-Memory Parallel Maximal Clique Enumeration}

\author{\IEEEauthorblockN{Apurba Das}
\IEEEauthorblockA{
Iowa State University\\
adas@iastate.edu}
\and
\IEEEauthorblockN{Seyed-Vahid Sanei-Mehri}
\IEEEauthorblockA{
Iowa State University\\
vas@iastate.edu}
\and
\IEEEauthorblockN{Srikanta Tirthapura}
\IEEEauthorblockA{
Iowa State University\\
snt@iastate.edu}
}

\maketitle

\begin{abstract}
We present shared-memory parallel methods for Maximal Clique Enumeration (MCE) from a graph. MCE is a fundamental and well-studied graph analytics task, and is a widely used primitive for identifying dense structures in a graph. Due to its computationally intensive nature, parallel methods are imperative for dealing with large graphs. However, surprisingly, there do not yet exist scalable and parallel methods for MCE on a shared-memory parallel machine. In this work, we present efficient shared-memory parallel algorithms for MCE, with the following properties: (1)~the parallel algorithms are provably work-efficient relative to a state-of-the-art sequential algorithm (2)~the algorithms have a provably small parallel depth, showing that they can scale to a large number of processors, and (3)~our implementations on a multicore machine shows a good speedup and scaling behavior with increasing number of cores, and are substantially faster than prior shared-memory parallel algorithms for MCE.
\end{abstract}

\section{Introduction}
\label{sec:intro}

We study the problem of Maximal Clique Enumeration (MCE) from a graph, which requires to enumerate all cliques (complete subgraphs) in the graph that are maximal. A clique $C$ in a graph $G = (V,E)$ is a dense subgraph such that every pair of vertices in $C$ are directly connected by an edge. A clique $C$ is said to be maximal when there is no clique $C'$ such that $C$ is a proper subgraph of $C'$. Maximal cliques are perhaps the most fundamental dense subgraphs, and MCE has been widely used in diverse research areas, such as clustering and community detection in social and biological networks~\cite{PD+05} and in genomics~\cite{RWY07}. It has also applications in finding common substructures in chemical compounds~\cite{KA+14}, mining from biological data~\cite{GA+93, HO+03, MBR04, CC05, JB06, ZP+08}, and inference from graphical models~\cite{KF09}.



MCE is a computationally hard problem since it is harder than the problem of finding the {\em maximum} clique, which is a classical NP-hard combinatorial problem. The computational cost of enumerating maximal cliques can be higher than the cost of finding the maximum clique, since the output size (set of all maximal cliques) may itself be very large, in the worst case. In particular, Moon and Moser~\cite{MM65} showed that a graph on $n$ vertices can have as many as $3^{n/3}$ maximal cliques, which is proven to be a tight bound. Real-world networks typically do not have cliques of such high complexity and it is possible to enumerate maximal cliques from large graphs. The literature is rich on sequential algorithms for MCE. Bron and Kerbosch~\cite{BK73} introduced a backtracking search method to enumerate maximal cliques. Tomita et. al \cite{TTT06} used the idea of ``pivoting'' in the backtracking search, which led to a significant improvement in the runtime. This has been followed up by further work such as due to Eppstein et al.~\cite{ELS10}, who used a degeneracy-based vertex ordering scheme on top of the pivot selection strategy.


Sequential approaches to MCE can lead to high runtimes on large graphs. Based on our experiments, a real-world network {\tt orkut} with approximately $3$ million vertices, $117$ million edges requires approximately $8$ hours to enumerate all maximal cliques using an efficient sequential algorithm due to Tomita et al.~\cite{TTT06}. Graphs that are larger and/or more complex cannot be handled by sequential algorithms with a reasonable turnaround time, and the high computational complexity of MCE calls for parallel methods.

In this work, we consider shared memory parallel methods for MCE. In the shared memory model, the input graph can reside within globally shared memory, and multiple threads can work in parallel on enumerating maximal cliques. Shared memory parallelism is attractive today since machines with tens to hundreds of cores and hundreds of Gigabytes of shared memory are readily available. The advantage of using shared memory approach over a distributed memory approach are: (1)~Unlike distributed memory, it is not necessary to divide the graph into subgraphs and communicate the subgraphs among processors. In shared memory, different threads can work with a single shared copy of the graph (2)~Subproblems generated during MCE are often irregular, and it is hard to predict which subproblems are small and which are large, while initially dividing the problem into subproblems. With a shared memory method, it is easy to further subdivide subproblems and process them in parallel. With a distributed memory method, handling such irregularly sized subproblems in a load-balanced manner requires greater coordination and is more complex.

Prior works on parallel MCE have largely focused on distributed memory algorithms~\cite{WY+09, SS+09, LGG10, SMT+15, XCF16}. There are a few works on shared-memory parallel algorithms~\cite{ZA+05, DW+09, LP+17}. However, these algorithms do not scale to larger graphs due to memory or computational bottlenecks -- either the algorithms miss out significant pruning opportunities as in~\cite{DW+09} or they need to generate a large number of non-maximal cliques as in~\cite{ZA+05, LP+17}. 

\subsection{Our Contributions}
We design shared-memory parallel algorithms for enumerating all maximal cliques in a simple graph. Our contributions are as follows: 

\textbf{Theoretically Efficient Parallel Algorithm: } We present a shared-memory parallel algorithm $\partomita$ that takes as input a graph $G$ and enumerates all maximal cliques in $G$. $\partomita$ is an efficient parallelization of the algorithm due to Tomita et al.~\cite{TTT06}. Our analysis of $\partomita$ using a work-depth model~\cite{S17} of computation shows that it is work-efficient when compared with~\cite{TTT06} and has a low parallel depth. To our knowledge, this is the first shared memory parallel algorithm for MCE with such provable properties.

\textbf{Optimized Parallel Algorithm:} We present the following ideas to further improve the practical performance of $\partomita$, leading to Algorithm~$\parmce$. First, instead of starting with a single task that spawns recursive subtasks as it proceeds, which leads to a lack of parallelism at the top level of recursion, we start with multiple parallel subtasks. To achieve this, we consider per-vertex parallelization, where a separate subproblem is created for each vertex and the different subproblems are processed in parallel. Each subproblem is required to enumerate cliques that contain the assigned vertex, where care is taken to prevent overlap between subproblems, and to balance the load between subproblems. 
Each per-vertex subproblem is further processed in parallel using $\partomita$. This additional (recursive) level of parallelism is useful since the different per-vertex subproblems may have significantly different computational costs, having each run as a separate sequential task may lead to uneven load balance. To further address load balance, we consider different methods for ranking the vertices, so that the ranking functions can be used in creating subproblems that are balanced as much as possible. For ranking the vertices, we use metrics such as degree, triangle count, and degeneracy number of the vertices.

\textbf{Experimental Evaluation: }We experimentally evaluate our algorithm and show that $\parmce$ is \textbf{15x-31x} faster than an efficient sequential algorithm (due to Tomita et al.~\cite{TTT06}) on a multicore machine with $32$ physical cores and $256$G RAM. For example, on the {\tt orkut} network with around $3$M vertices, $117$M edges, and $2$B maximal cliques \footnote{M and B stand for million and billion respectively.}, \partomita achieves a \textbf{14x} parallel speedup over the sequential algorithm, and the optimized $\parmce$ achieves a \textbf{16x} speedup. In contrast, prior shared memory parallel algorithms for MCE~\cite{ZA+05,DW+09,LP+17} failed to handle the input graphs that we considered, and either ran out of memory (\cite{ZA+05,LP+17}) or did not complete in 5 hours (\cite{DW+09}).



\textbf{Roadmap.} The rest of the paper is organized as follows. We present preliminaries in Section~\ref{sec:prelims}, followed by a description of the algorithm and analysis in Section~\ref{sec:algo}, an experimental evaluation in Section~\ref{sec:exp}, and conclusions in Section~\ref{sec:conclude}.

\section{Related Work}
\label{sec:related}
Maximal Clique Enumeration (MCE) from a graph is a fundamental problem that has been extensively studied for more than two decades, and there are multiple prior works on sequential and parallel algorithms. We first discuss sequential algorithms for MCE, followed by parallel algorithms.

\textbf{Sequential MCE:} Bron and Kerbosch~\cite{BK73} presented an algorithm for MCE based on depth-first-search. Following their work, a number of algorithms have been presented ~\cite{TI+77,CN85,TTT06,Koch01,MU04,CK08,ELS10}. The algorithm of Tomita et al.~\cite{TTT06} has a worst-case time complexity $O(3^{\frac{n}{3}})$ for an $n$ vertex graph, which is optimal in the worst-case, since the size of the output can be as large as $O(3^{\frac{n}{3}})$~\cite{MM65}. Eppstein et al.~\cite{ELS10,ES11} present an algorithm for sparse graphs whose complexity can be parameterized by the degeneracy of the graph, a measure of graph sparsity.

Another approach to MCE is a class of ``output-sensitive" algorithms whose time complexity for enumerating maximal cliques is a function of the size of the output. There exist many such output-sensitive algorithms for MCE, including~\cite{CN85,MU04,TI+77}, which can be viewed as instances of a general paradigm called ``reverse-search''~\cite{AF93}. The output-sensitive algorithm due to Makino and Uno~\cite{MU04} provides the best theoretical worst-case time complexity among output-sensitive algorithms. In terms of practical performance, the best output-sensitive algorithms~\cite{CN85,MU04} are not as efficient as the best depth-first-search based algorithms such as~\cite{TTT06,ELS10}. Other sequential methods for MCE include algorithms due to Kose et al.~\cite{KW+01}, Johnson et al.~\cite{JYP88}, and Cheng et al.~\cite{CK+11}. There have also been works on maintaining maximal cliques in a dynamic graph~\cite{DST16,S04,OV10}.

\textbf{Parallel MCE: }There are multiple prior works on parallel algorithms for MCE ~\cite{ZA+05,DW+06,WY+09,SS+09,LGG10,CZ+12,SMT+15,XCF16}. We first discuss shared memory algorithms and then distributed memory algorithms.

Zhang et al.~\cite{ZA+05} presented a shared memory parallel algorithm based on the sequential algorithm due to Kose et al.~\cite{KW+01}. This algorithm computes maximal cliques in an iterative manner, and in each iteration, it maintains a set of cliques that are not necessarily maximal and for each such clique, maintains the set of vertices that can be added to form larger cliques. This algorithm does not provide a theoretical guarantee on the runtime and suffers for large memory requirement. Du et al.~\cite{DW+06} present a output-sensitive shared-memory parallel algorithm for MCE, but their algorithm suffers from poor load balancing as also pointed out by Schmidt et al.~\cite{SS+09}. Lessley et al.~\cite{LP+17} present a shared memory parallel algorithm that generates maximal cliques using an iterative method, where in each iteration, cliques of size $(k-1)$ are extended to cliques of size $k$.  The algorithm of~\cite{LP+17} is memory-intensive, since it needs to store a number of intermediate non-maximal cliques in each iteration. Note that the number of non-maximal cliques may be far higher than the number of maximal cliques that are finally emitted, and a number of distinct non-maximal cliques may finally lead to a single maximal clique. In the extreme case, a complete graph on $n$ vertices has $(2^n-1)$ non-maximal cliques, and only a single maximal clique. We present a comparison of our algorithm with~\cite{LP+17,ZA+05,DW+06} in later sections. 

Distributed memory parallel algorithms for MCE include works due to Wu et al.~\cite{WY+09}, designed for the MapReduce framework, Lu et al.~\cite{LGG10}, which is based on the sequential algorithm due to Tsukiyama et al.~\cite{TI+77}, Xu et al.~\cite{XCF16}, and Svendsen et al.~\cite{SMT+15}. 

Other works on parallel algorithms for enumerating dense subgraphs from a massive graph include parallel algorithms for enumerating $k$-cores~\cite{MD+13,DDZ14,KM17,SSP17}, $k$-trusses~\cite{SSP17,KM17-2,KM17-3}, nuclei~\cite{SSP17}, and distributed memory algorithms for enumerating bicliques~\cite{MT17}.


\section{Preliminaries}
\label{sec:prelims}
We consider a simple undirected graph without self loops or multiple edges. For graph $G$, let $V(G)$ denote the set of vertices in $G$ and $E(G)$ denote the set of edges in $G$. Let $n$ denote the size of $V(G)$, and $m$ denote the size of $E(G)$. For vertex $u \in V(G)$, let $\Gamma_G(u)$ denote the set of vertices adjacent to $u$ in $G$. When the graph $G$ is clear from the context, we use $\Gamma(u)$ to mean $\Gamma_G(u)$. Let $\cliques(G)$ denote the set of all maximal cliques in $G$. 

\textbf{Sequential Algorithm $\tomita$: } The algorithm due to Tomita, Tanaka, and Takahashi.~\cite{TTT06}, which we call $\tomita$, is a recursive backtracking-based algorithm for enumerating all maximal cliques in an undirected graph, with a worst-case time complexity of $O(3^{n/3})$ where $n$ is the number of vertices in the graph. In practice, this is one of the most efficient sequential algorithms for MCE. Since we use $\tomita$ as a subroutine in our parallel algorithms, we present a short description here.

In any recursive call, $\tomita$ maintains three disjoint sets of vertices $K$, $\cand$, and $\fini$ where $K$ is a candidate clique to be extended, $\cand$ is the set of vertices that can be used to extend $K$, and $\fini$ is the set of vertices that are adjacent to $K$, but need not be used to extend $K$ (these are being explored along other search paths). Each recursive call iterates over vertices from $\cand$ and in each iteration, a vertex $q \in \cand$ is added to $K$ and a new recursive call is made with parameters $K\cup\{q\}$, $\cand_q$, and $\fini_q$ for generating all maximal cliques of $G$ that extend $K\cup\{q\}$ but do not contain any vertices from $\fini_q$. The sets $\cand_q$ and $\fini_q$ can only contain vertices that are adjacent to all vertices in $K\cup \{q\}$. The clique $K$ is a maximal clique when both $\cand$ and $\fini$ are empty.

The ingredient that makes $\tomita$ different from the algorithm due to Bron and Kerbosch~\cite{BK73} is the use of a ``pivot'' where a vertex $u\in \cand\cup\fini$ is selected that maximizes $\vert \cand \cap \Gamma(u) \vert$. Once the pivot $u$ is computed, it is sufficient to iterate over all the vertices of $\cand \setminus \Gamma(u)$, instead of iterating over all vertices of $\cand$. The pseudo code of $\tomita$ is presented in Algorithm~\ref{algo:tomita}. For the initial call, $K$ and $\fini$ are initialized to an empty set, $\cand$ is the set of all vertices of $G$. 


\begin{algorithm}
\DontPrintSemicolon
\caption{$\tomita(\mathcal{G},K,\cand,\fini)$}
\label{algo:tomita}
\KwIn{$\mathcal{G}$ - The input graph \\ \hspace{1cm} $K$ - a clique to extend, \\
$\cand$ - Set of vertices that can be used extend $K$, \\ 
$\fini$ - Set of vertices that have been used to extend $K$}
\KwOut{Set of all maximal cliques of $G$ containing $K$ and vertices from $\cand$ but not containing any vertex from $\fini$}
\If{$(\cand = \emptyset)$ \& $(\fini = \emptyset)$}{
	Output $K$ and return\;
}
$\pivot \gets (u \in \cand \cup \fini)$ such that $u$ maximizes the size of $\cand  \cap \Gamma_{\mathcal{G}}(u)$\;
$\ext \gets \cand - \Gamma_{\mathcal{G}}(\pivot)$\;
\For{$q \in$ \ext}{
	$K_q \gets K\cup\{q\}$\;
	$\cand_q \gets \cand\cap\Gamma_{\mathcal{G}}(q)$\;
	$\fini_q \gets \fini\cap\Gamma_{\mathcal{G}}(q)$\;
    $\cand \gets \cand-\{q\}$\;
	$\fini \gets \fini \cup\{q\}$\;
	$\tomita(\mathcal{G},K_q,\cand_q,\fini_q)$\;
}
\end{algorithm}

\textbf{Parallel Cost Model: } For analyzing our shared-memory parallel algorithms, we use the CRCW PRAM model~\cite{BM10}, which is a model of shared parallel computation that assumes concurrent reads and concurrent writes. Our parallel algorithm can also work in other related models of shared memory such as EREW PRAM (exclusive reads and exclusive writes), with a logarithmic factor increase in work as well as parallel depth. We measure the effectiveness of the parallel algorithm using the {\em work-depth} model~\cite{S17}. Here, the ``work'' of a parallel algorithm is equal to the total number of operations of the parallel algorithm, and the ``depth'' (also called the ``parallel time'' or the ``span'') is the longest chain of dependent computations in the algorithm. A parallel algorithm is said to be {\em work-efficient} if its total work is of the same order as the work due to the best sequential algorithm\footnote{Note that work-efficiency in the CRCW PRAM model does not imply work-efficiency in the EREW PRAM model}. We aim for work-efficient algorithms with a low depth, ideally poly-logarithmic in the size of the input. Using Brent's theorem~\cite{BM10}, it can be seen that a parallel algorithm on input size $n$ with a depth of $d$ can theoretically achieve $\Theta(p)$ speedup on $p$ processors as long as $p = O(n/d)$. 




\section{Parallel MCE Algorithms}
\label{sec:algo}
In this section, we present shared-memory parallel algorithms for MCE. We first describe a parallel algorithm $\partomita$ and an analysis of its theoretical properties, where $\partomita$ is a parallel version of the $\tomita$ algorithm. Then, we discuss practical bottlenecks in $\partomita$, leading us to another algorithm $\parmce$ with a better practical runtime performance. 
 
\subsection{Algorithm $\partomita$}
Our first algorithm $\partomita$ is a work-efficient parallelization of the sequential $\tomita$ algorithm. The two main components of $\tomita$ (Algorithm~\ref{algo:tomita}) are (1)~Selection of the pivot element (Line~$3$) and (2)~Sequential backtracking for extending candidate cliques until all maximal cliques are explored (Line~$5$ to Line~$11$). We discuss how to parallelize each of these steps.

\textbf{Parallel Pivot Selection: } Within a single recursive call of $\partomita$, the pivot element is computed in parallel using two steps, as described in $\parpivot$ (Algorithm~\ref{algo:pivot}). In the first step, the size of the intersection $\cand \cap \Gamma(u)$ is computed in parallel for each vertex $u \in \cand \cup \fini$. In the second step, the vertex with the maximum intersection size is selected. The parallel algorithm for selecting a pivot is presented in Algorithm~\ref{algo:pivot}.

\begin{lemma}
\label{lem:parpivot}
The total work of $\parpivot$ is $O(\sum_{w \in \cand \cup \fini}  (\min\{|\cand|, |\Gamma(w)|\})$, which is $O(n^2)$, and depth is $O(\log n)$.
\end{lemma}

\begin{IEEEproof}
If the sets $\cand$ and $\Gamma(w)$ are stored as hashsets, then for vertex $w$ the size $t_w = \vert\intersect(\cand, \Gamma(w))\vert$ can be computed sequentially in time $O(\min\{|\cand|, |\Gamma(w)|\})$ -- the intersection of two sets $S_1$ and $S_2$ can be found by considering the smaller set among the two, say $S_2$, and searching for its elements within the larger set, say $S_1$. It is possible to parallelize the computation of $\intersect(S_1,S_2)$ by executing the searches elements in $S_2$ in parallel, followed by counting the number of elements that lie in the intersection, which can also be done in parallel in a work-efficient manner using logarithmic depth. Since computing the maximum of a set of $n$ numbers can be accomplished using work $O(n)$ and depth $O(\log n)$, for vertex $w$, $t_w$ can be computed using work $O(\min\{|\cand|, |\Gamma(w)|\})$ and depth $O(\log n)$. Once the different $t_w$ are computed, $argmax(\{t_w : w \in \cand\cup\fini\})$ can be computed using additional work $\vert \cand \cup \fini \vert$ and depth $O(\log n)$. Hence, the total work of $\parpivot$ is $O(\sum_{w \in \cand \cup \fini}  (\min\{|\cand|, |\Gamma(w)|\})$. Since the size of $\cand$, $\fini$, and $\Gamma(w)$ are bounded by $n$, this is $O(n^2)$,  but typically much smaller.
\end{IEEEproof}

\begin{algorithm}
\DontPrintSemicolon
\caption{$\parpivot(\mathcal{G},K, \cand,\fini)$}
\label{algo:pivot}
\KwIn{ 
$K$ - a clique in $G$ that may be further extended \\
$\mathcal{G}$ - Input graph \\
$\cand$ - Set of vertices that may extend $K$ \\ 
$\fini$ - vertices that have been used to extend $K$}
\KwOut{
pivot vertex $u \in \cand\cup\fini$
}
\ForPar {$w \in \cand\cup\fini$}{
	In parallel, compute $t_w \gets |\intersect(\cand, \Gamma_{\mathcal{G}}(w))|$\;
}
In parallel, find $v \gets argmax(\{t_w : w \in \cand\cup\fini\})$\;
\Return{$v$}\;
\end{algorithm}

\textbf{Parallelization of Backtracking: } We first note that there is a sequential dependency among the different iterations within a recursive call of $\tomita$. In particular, the contents of the sets $\cand$ and $\fini$ in a given iteration are derived from the contents of $\cand$ and $\fini$ in the previous iteration. Such sequential dependence of updates of $\cand$ and $\fini$ restricts us from calling the recursive $\tomita$ for different vertices of $\ext$ in parallel. To remove this dependency, we adopt a different view of $\tomita$ which enables us to make the recursive calls in parallel. The elements of $\ext$, the vertices to be considered for extending a maximal clique, are arranged in a predefined total order. Then, we unroll the loop and explicitly compute the parameters \cand and \fini for recursive calls. 

Suppose $\langle v_1, v_2, ..., v_{\kappa} \rangle$ is the order of vertices in $\ext$ to be processed in sequence. Each vertex $v_i \in \ext$, once added to $K$, should be removed from further consideration from $\cand$. To ensure this, instead of incrementally updating $\cand$ and $\fini$ with $v_i$ as in $\tomita$, in $\partomita$, we explicitly remove vertices $v_1, v_2, ..., v_{i-1}$ from $\cand$ and add them to $\fini$, before making the recursive calls. This way, the parameters of the $i$th iteration are computed independently of prior iterations. 




\begin{algorithm}
\DontPrintSemicolon
\caption{$\partomita(\mathcal{G},K,\cand,\fini)$}
\label{algo:partomita}
\KwIn{$\mathcal{G}$ - The input graph \\ \hspace{1cm} $K$ - a non-maximal clique to extend \\ 
$\cand$ - Set of vertices that may extend $K$ \\ 
$\fini$ - vertices that have been used to extend $K$}
\KwOut{Set of all maximal cliques of $G$ containing $K$ and vertices from $\cand$ but not containing any vertex from $\fini$}
\If{$(\cand = \emptyset)$ \& $(\fini = \emptyset)$}{
	Output $K$ and return\;
}
$\pivot \gets \parpivot(\mathcal{G},\cand,\fini)$\;
$\ext[1..\kappa] \gets \cand - \Gamma_{\mathcal{G}}(\pivot)$ \tcp{in parallel} \;
\ForPar{$i \in$ $[1..\kappa]$}{
	$q\gets\ext[i]$\;
	$K_q \gets K\cup\{q\}$\;
	$\cand_q \gets \intersect(\cand\setminus\ext[1..i-1], \Gamma_{\mathcal{G}}(q))$\;
	$\fini_q \gets \intersect(\fini\cup\ext[1..i-1], \Gamma_{\mathcal{G}}(q))$\;
	$\partomita(\mathcal{G},K_q,\cand_q,\fini_q)$\;
}
\end{algorithm}

Next, we present an analysis of the total work and depth of $\parpivot$ and $\partomita$ algorithms.


\begin{lemma}
\label{lem:partomita}
Total work of $\partomita$ (Algorithm~\ref{algo:partomita}) is $O(3^{n/3})$ and depth is $O(M\log{n})$ where $n$ is the number of vertices in the graph and $M$ is the size of maximum clique in $G$.
\end{lemma} 

\begin{IEEEproof}
First, we analyze the total work. Note that the computational tasks in $\partomita$ is different from $\tomita$ at Line~9 and Line~10 of $\partomita$ where at an iteration $i$, we remove all vertices $\{v_1, v_2, ..., v_{i-1}\}$ from $\cand$ and add all these vertices to $\fini$ as opposed to the removal of a single vertex $v_{i-1}$ from $\cand$ and addition of that vertex to $\fini$ as in $\tomita$ (Line~9 and Line~10 of Algorithm~\ref{algo:tomita}). Therefore, in $\partomita$, additional $O(n)$ work is required due to independent computations of $\cand_q$ and $\fini_q$. The total work, excluding the call to $\parpivot$ is $O(n^2)$. Adding up the work of $\parpivot$, which requires $O(n^2)$ work, requires $O(n^2)$ total work for each single call of $\partomita$ excluding further recursive calls (Algorithm~\ref{algo:partomita}, Line 11), which is same as in original sequential algorithm $\tomita$ (Section 4, \cite{TTT06}). Hence, using Lemma~2 and Theorem~3 of \cite{TTT06}, we infer that the total work of $\partomita$ is the same as the sequential algorithm $\tomita$ and is bounded by $O(3^{n/3})$. 

Next we analyze the depth of the algorithm. The depth of $\partomita$ consists of the (sum of the) following components: (1)~Depth of $\parpivot$, (2)~Depth of computation of $\ext$, (3)~Maximum depth of an iteration in the for loop from Line~6 to Line~11. According to Lemma~\ref{lem:parpivot}, the depth of $\parpivot$ is $O(\log{n})$. The depth of computing $\ext$ is $O(\log n)$ because it takes $O(1)$ time to check whether an element in $\cand$ is in the neighborhood of $\pivot$ by doing a set membership check on the set of vertices that are adjacent to $\pivot$. Similarly, the depth of computing $\cand_q$ and $\fini_q$ at Line~8 and Line~9 are $O(\log n)$ each. The remaining is the depth of the call of $\partomita$ at Line~10. Observe that the recursive call of $\partomita$ continues until there is no further vertex to add for expanding $K$, and this depth can be at most the size of the maximum clique which is $M$ because, at each recursive call of $\partomita$ the size of $K$ is increased by $1$. Thus, the overall depth of $\partomita$ is $O(M\log{n})$. 
\end{IEEEproof}

\begin{corollary}
\label{cor:partomita}
Using $P$ parallel processors that share memory, $\partomita$ (Algorithm~\ref{algo:partomita}) is a parallel algorithm for MCE, and can achieve a worst case parallel time of $O\left(\frac{3^{n/3}}{M \log n} + P\right)$ using $P$ parallel processors. This is work-efficient as long as $P = O(\frac{3^{n/3}}{M \log n})$, and also work-optimal.
\end{corollary}

\begin{IEEEproof}
The parallel time follows from using Brent's theorem~\cite{BM10}, which states that the parallel time using $P$ processors is $O(w/d + P)$, where $w$ and $d$ are the work and the depth of the algorithm respectively. If the number of processors $P= O\left(\frac{3^{n/3}}{M \log n}\right)$, then using Lemma~\ref{lem:partomita} the parallel time is $O\left(\max\{\frac{3^{n/3}}P , {M \log n}\}\right) = O\left(\frac{3^{n/3}}{P}\right)$. The total work across all processors is $O(3^{n/3})$, which is worst-case optimal, since the size of the output can be as large as $3^{n/3}$ maximal cliques (Moon and Moser~\cite{MM65}). 
\end{IEEEproof}

\subsection{Algorithm $\parmce$}
While $\partomita$ is a theoretically work-efficient parallel algorithm, we note that it is not that efficient in practice. One of the reasons is the implementation of $\parpivot$. While the worst case work complexity of $\parpivot$ matches that of the pivoting routine in $\tomita$, in practice, it may have a higher overhead, since the pivoting routine in $\tomita$ may take time less than $O(n^2)$. This can cause $\partomita$ to have greater work than $\tomita$, resulting in a lower speedup than the theoretically expected one.


We set out to improve on this to derive a more efficient parallel implementation through a more selective use of $\parpivot$ in that the cost of pivoting can be reduced by carefully choosing many pivots in parallel instead of a single pivot element as in $\partomita$ at the beginning of the algorithm. We first note that the cost of $\parpivot$ is the highest during the iteration when the parameter $K$ (clique so far) is empty. During this iteration, the set of vertices still to be considered, $\cand \cup \fini$, can be high, as large as the number of vertices in the graph. To improve upon this, we can perform the first few steps of pivoting, when $K$ is empty, using a sequential algorithm. Once the set $K$ has at least one element in it, the number of the vertices in $\cand \cup \fini$ still to be considered, drops down to no more than the size of the intersection of neighborhoods of all vertices in $K$, which is typically a number much smaller than the number of vertices in the graph (it is smaller than the smallest degree of a vertex in $K$). Problem instances with $K$ set to a single vertex can be seen as subproblems and on each of these subproblems, the overhead of $\parpivot$ is much smaller since the number of vertices that have to be dealt with is also much smaller.

Based on this observation, we present a parallel algorithm $\parmce$ that works as follows. The algorithm can be viewed as considering for each vertex $v \in V(G)$, a subgraph $G_v$ that is induced by the vertex $v$ and its neighborhood $\Gamma_G(v)$. It enumerates all maximal cliques from each subgraph $G_v$ in parallel using $\partomita$. While processing subproblem $G_v$, it is important to not enumerate maximal cliques that are being enumerated elsewhere, in other subproblems. To handle this, the algorithm considers a specific ordering of all vertices in $V$ such that $v$ is the least ranked vertex in each maximal clique enumerated from $G_v$. The subgraphs $G_v$ for each vertex $v$ are handled in parallel -- these subgraphs need not be processed in any particular order. However, the ordering allows us to populate the $\cand$ and $\fini$ sets accordingly, so that each maximal clique is enumerated in exactly one subproblem. The order in which the vertices are considered is defined by a ``rank'' function \textbf{rank}, which indicates the position of a vertex in the total order. The specific ordering that is used influences the total work of the algorithm, as well as the load balance of the parallel implementation.


\textbf{Load Balancing:} Observe that the sizes of the subgraphs $G_v$ may vary widely because of two reasons: (1)~the subgraphs themselves may be of different sizes, depending on the vertex degrees, and (2)~the number of maximal cliques and the sizes of the maximal cliques containing $v$ can vary widely from one vertex to another. Clearly, the subproblems that deal with a large number of maximal cliques or maximal cliques of a large size are more expensive than others.

In order to maintain the size of the subproblems approximately balanced, we use an idea from PECO~\cite{SMT+15}, where we choose the rank function on the vertices in such a way that for any two vertices $v$ and $w$, \textbf{rank}($v$) $>$ \textbf{rank}($w$) if the complexity of enumerating maximal cliques from $G_v$ is higher than the complexity of enumerating maximal cliques from $G_w$. By giving a higher rank to $v$ than $w$, we are decreasing the complexity of the subproblem $G_v$, since the subproblem at $G_v$ need not enumerate maximal cliques that involve any vertex whose rank is less than $v$. Hence, the higher the rank of vertex $v$, the lower is its ``share'' (of maximal cliques it belongs to) of maximal cliques in $G_v$. We use this idea for approximately balancing the workload across subproblems. The additional enhancements in $\parmce$, when compared with the idea from PECO are as follows: (1)~In PECO the algorithm is designed for distributed memory so that the subgraphs and subproblems have to be explicitly copied across the network, and (2)~In $\parmce$, the vertex specific subproblem, dealing with $G_v$ is itself handled through a parallel algorithm, $\partomita$. However, in PECO, the subproblem for each vertex was handled through a sequential algorithm.


Note that it is computationally expensive to accurately count the number of maximal cliques within $G_v$, and hence it is not possible to compute the rank of each vertex exactly according to the complexity of handling $G_v$. Instead, we estimate the complexity of handling $G_v$ using some easy-to-evaluate metrics on the subgraphs. In particular, we consider the following:


\begin{itemize}
\item 
\textbf{Degree Based Ranking: } For vertex $v$, define $\rank(v) = (d(v), id(v))$ where $d(v)$ and $id(v)$ are degree and identifier of $v$ respectively. For two vertices $v$ and $w$, $\rank(v) > \rank(w)$ if $d(v) > d(w)$ or $d(v) = d(w)$ and $id(v) > id(w)$; $rank(v) < rank(w)$ otherwise. 

\item
\textbf{Triangle Count Based Ranking: } For vertex $v$, define $\rank(v) = (t(v), id(v))$ where $t(v)$ is the number of triangles containing vertex $v$. This is more expensive to compute than degree based ranking, but may yield a better estimate of the complexity of maximal cliques within $G_v$.

\item
\textbf{Degeneracy Based Ranking~\cite{ELS10}: } For a vertex $v$, define $\rank(v) = (degen(v), id(v))$ where $degen(v)$ is the degeneracy of a vertex $v$. A vertex $v$ has degeneracy number $k$ when it belongs to a $k$-core but no $(k+1)$-core where a $k$-core is a maximal induced subgraph with minimum degree of each vertex $k$ in that subgraph. A computational overhead of using this ranking is due to computing the degeneracy of the vertices which takes $O(n+m)$ time where $n$ is the number of vertices and $m$ is the number of edges.
\end{itemize}

The different implementations of $\parmce$ using degree, triangle, and degeneracy rankings are called as $\parmcedegree$, $\parmcetriangle$, $\parmcedegen$ respectively. 


\begin{algorithm}
\DontPrintSemicolon
\caption{$\parmce(\mathcal{G})$}
\label{algo:parmce}
\KwIn{$\mathcal{G}$ - The input graph}
\KwOut{$\cliques(\mathcal{G})$ - set of all maximal cliques of $\mathcal{G}$}
\ForPar {$v \in V(\mathcal{G})$}{
	Create $\mathcal{G}_v$, the subgraph of $\mathcal{G}$ induced by $\Gamma_{\mathcal{G}}(v)\cup\{v\}$\;
	$K \gets \{v\}$\;
    $\cand \gets \phi$\;
    $\fini \gets \phi$\;
    \ForPar {$w \in \Gamma_{\mathcal{G}}(v)$}{
    	\If{$rank(w) > rank(v)$}{
        	$\cand \gets \cand\cup\{w\}$\;
        }
        \Else {
        	$\fini \gets \fini\cup\{w\}$\;
        }
    }
    $\partomita(\mathcal{G}_v, K, \cand, \fini)$\;
}
\end{algorithm}
\section{Experiments}
\label{sec:exp}
In this section, we present results from an experimental evaluation of the performance of parallel algorithms for MCE. For our experiments, we used an Intel Xeon (R) CPU on a Compute Engine in the Google Cloud Platform, with $32$ physical cores and $256$ GB RAM. We implement all algorithms using java 1.8 with a maximum of $100$ GB heap memory for the JVM. 

\subsection{Datasets}
We use large real world network datasets from KONECT~\cite{konnect13}, SNAP~\cite{JA14}, and Network Repository~\cite{RN15}. Table~\ref{graph:summary} contains a summary of the datasets. All networks, used in our experiments, were undirected graphs. Self-loops are removed, and if the input graph was directed, we ignored the direction on the edges to derive an undirected graph.

\begin{table*}[htp!]
\centering
\begin{tabular}{| l | r | r | r | r | r |}
\toprule
\textbf{Dataset} & $|V|$ & $|E|$ & \# maximal cliques & avg. size of a maximal clique & size of the maximum clique\\
\midrule
{\tt dblp-coauthors} & \num{540486} & \num{15245729} & \num{139340} & 11 & 337\\
\hline
{\tt orkut} & \num{3072441} & \num{117184899} & \num{2270456447} & 20 & 51\\
\hline
{\tt as-skitter} & \num{1696415} & \num{11095298} & \num{37322355} & 19 & 67\\
\hline
{\tt wiki-talk} & \num{2394385} & \num{4659565} & \num{86333306} & 13 & 26\\
\bottomrule
\end{tabular}
\caption{\textbf{Undirected graphs used for evaluation, and their properties.}}
\label{graph:summary}
\end{table*}

\subsection{Implementation of the algorithms}
In our implementation of $\partomita$ and $\parmce$, we implement a parallel \texttt{For} loop using the primitive {\tt parallelStream()} provided by java 1.8. For computing the intersection of two sets as is required for computing $\pivot$ and updating $\cand$ and $\fini$ in Algorithm \ref{algo:partomita}, we perform a sequential implementation. This is because the sizes of the sets $\cand$ and $\fini$ are typically not large so that we can benefit from the use of parallelism. For the garbage collection in java we use flag {\tt -XX:+UseParallelGC} so that parallel garbage collection is run by JVM whenever required. 

To compare with prior works in maximal clique enumeration, we implemented some of them \cite{TTT06,ELS10,DW+09,SMT+15,ZA+05} in Java, except the sequential algorithm $\greedybb$~\cite{SAS18}, and the parallel algorithm $\hashing$~\cite{LP+17}, for which we used the executables provided by the authors (code written in C++). See Subsection \ref{sec:exp-prior} for more details.



We call our implementation of $\parmce$ using degree based vertex ordering as $\parmcedegree$, using degeneracy based vertex ordering as $\parmcedegen$, and using triangle count based vertex ordering as $\parmcetriangle$. We compute the degeneracy number and triangle count for each vertex using sequential procedures. While the computation of per-vertex triangle counts and the degeneracy ordering could be potentially parallelized, implementing a parallel method to rank vertices based on their degeneracy number or triangle count is in itself a non-trivial task. We decided not to parallelize these routines since the degeneracy- and triangle-based ordering did not yield significant benefits when compared with degree-based ordering, where as degree-based ordering is trivially available, without any additional computation. 

We assume that the entire graph is stored in available in shared global memory. The runtime of $\parmce$ consists of (1)~the time required to rank vertices of the graph based on the ranking metric used in the algorithm, i.e. degree, degeneracy number, or triangle count of vertices and (2)~the time required to enumerate all maximal cliques. For $\parmcedegen$ and $\parmcetriangle$ algorithms, the runtime of ranking  is also reported. Figures~\ref{fig:parallel-speedup-linear-scale} and~\ref{fig:runtimes-per-thread} show the parallel speedup (with respect to the runtime of $\tomita$) and and the total computation time of $\parmce$ using different vertex ordering strategies, respectively. Table~\ref{result:runtime-splitup} shows the breakdown of the runtime into time for ordering and the time for clique enumeration. 
 




\subsection{Performance of Parallel Clique Enumeration Algorithms}

\begin{table*}[htp!]
\centering
\begin{tabular}{| l | r | r | r | r | r |}
\toprule
\textbf{DataSet} & \tomita & \partomita & \parmcedegree & \parmcedegen & \parmcetriangle \\
\midrule
{\tt dblp-coauthors} & 356 & 28 & 14 & 21.4 & 152.2\\
\hline
{\tt orkut} & \num{26407} & \num{1886} & \num{1362} & \num{2141.1} & \num{2278}\\
\hline
{\tt as-skitter} &  807 & 60 & 45 & 71.9 & 85.6\\
\hline
{\tt wiki-talk} & \num{1022} & \num{85} & 62 & 70.1 & 89.2\\
\bottomrule
\end{tabular}
\caption{\textbf{Comparison of total computation time (in sec.) of $\parmce$ (with degree based, degeneracy based, and triangle count based vertex ordering) and computation time $\partomita$ (with $32$ threads) with $\tomita$.}}
\label{result:parallel-runtimes}
\end{table*}

\begin{figure*}[htp!]
	\begin{tabular}{c}
		\centering
		\includegraphics[width=0.75\textwidth]{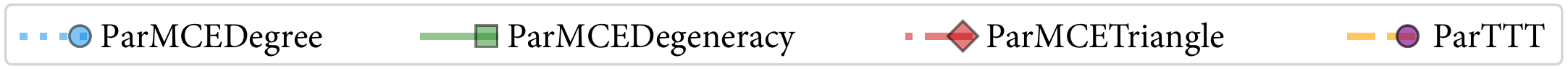}
		\vspace{-0.2ex}
	\end{tabular}
	\centering
	\begin{tabular}{cccc}
		\includegraphics[width=.23\textwidth]{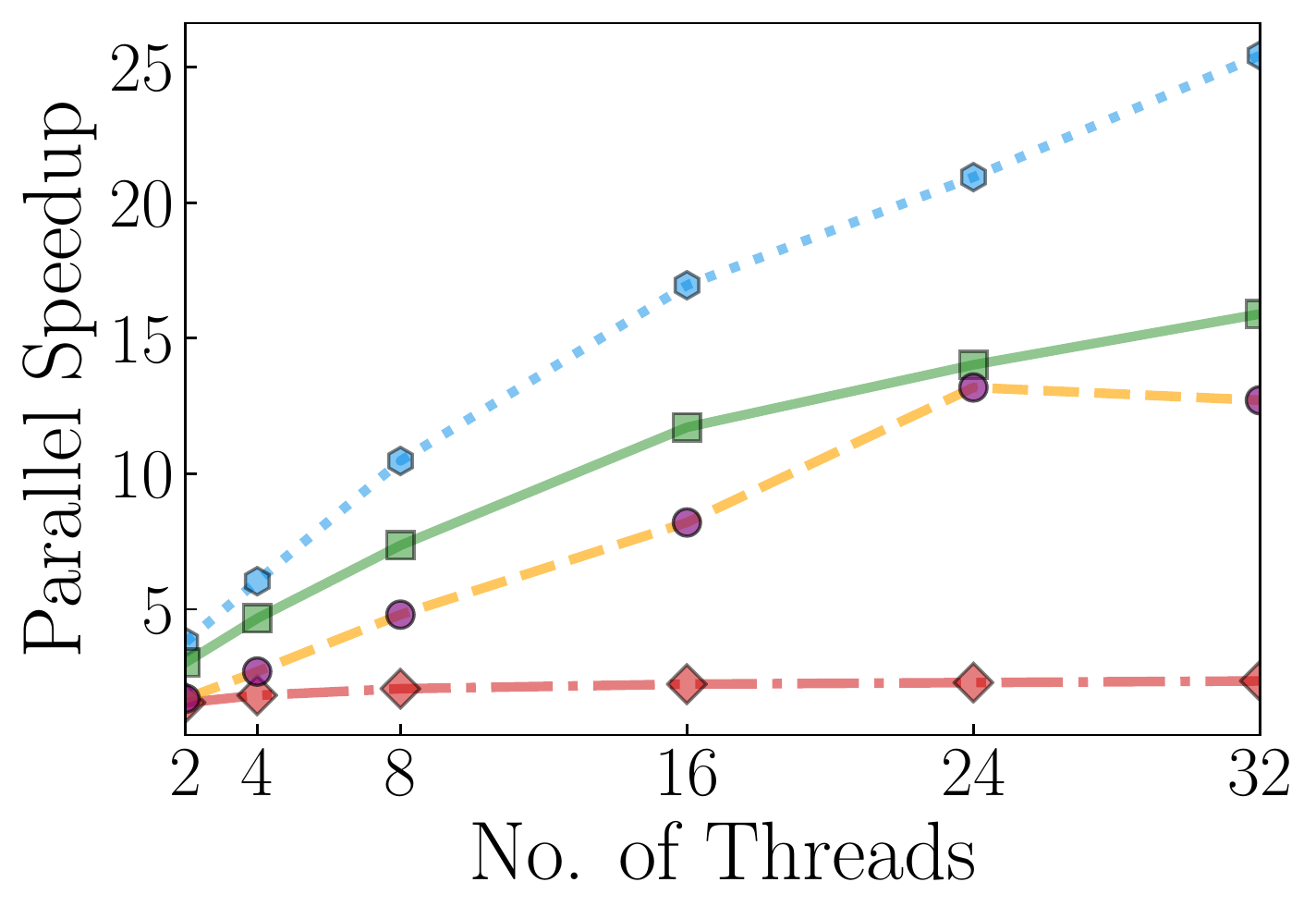} &
		\includegraphics[width=.23\textwidth]{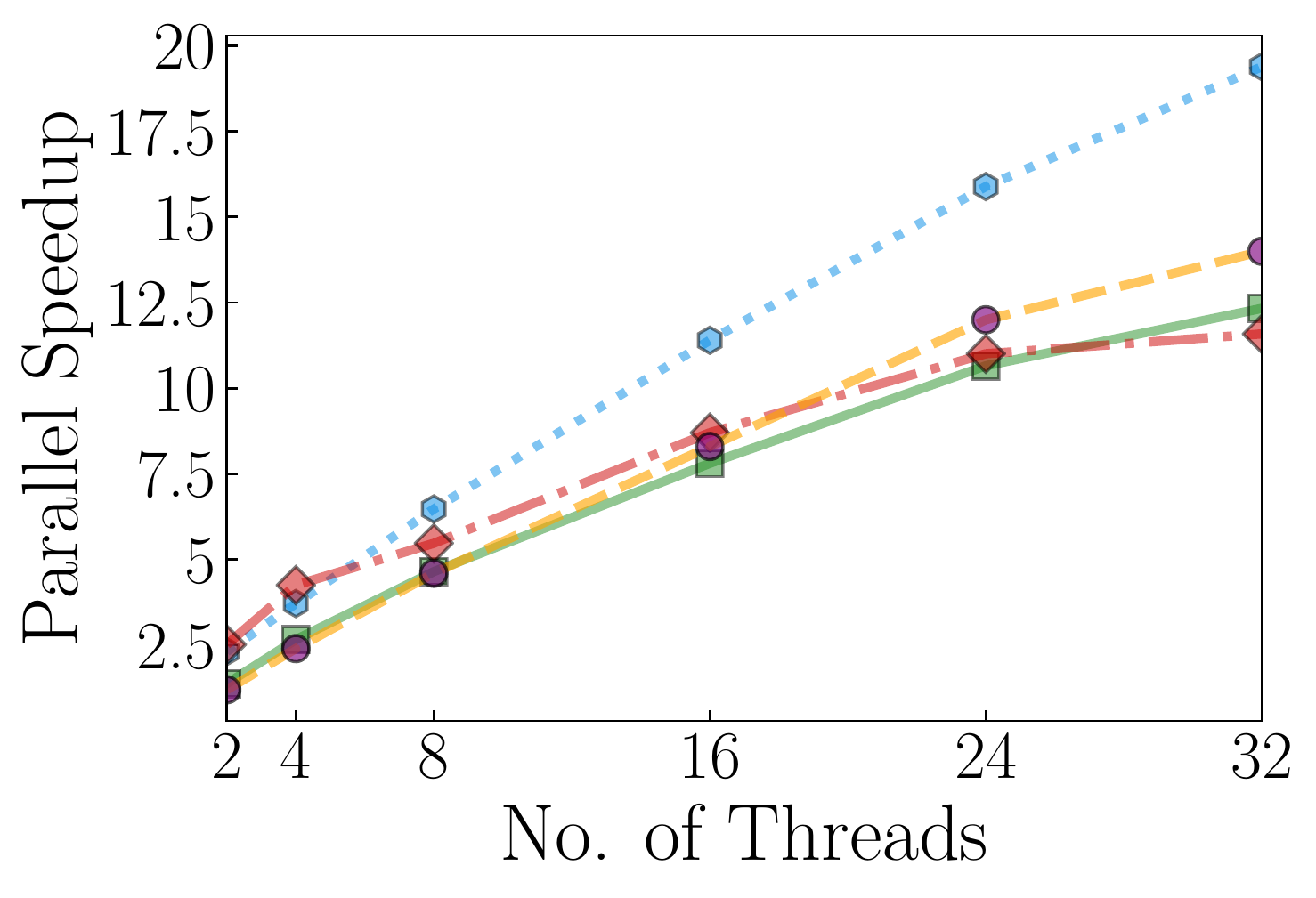} &
		\includegraphics[width=.23\textwidth]{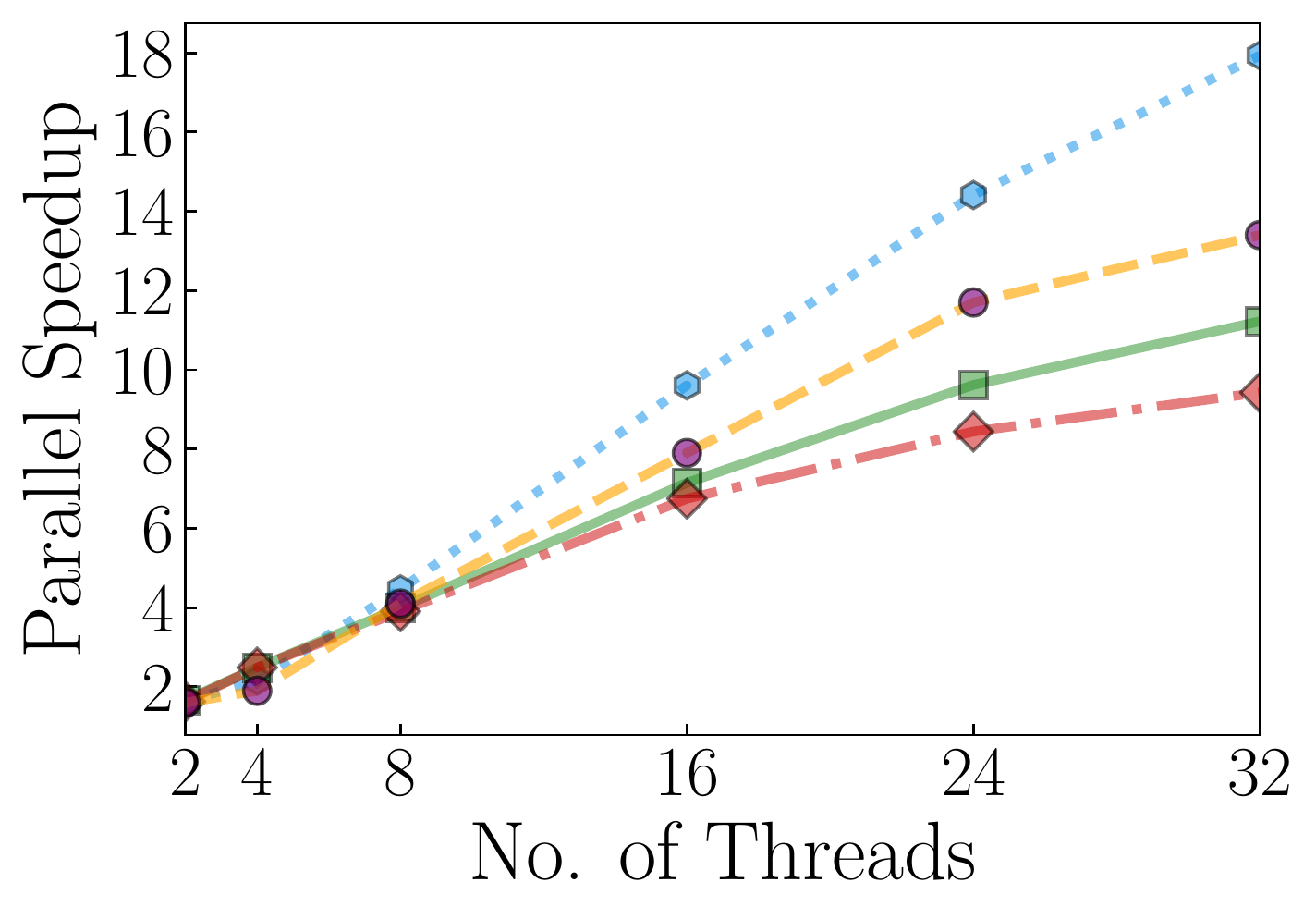} &
		\includegraphics[width=.23\textwidth]{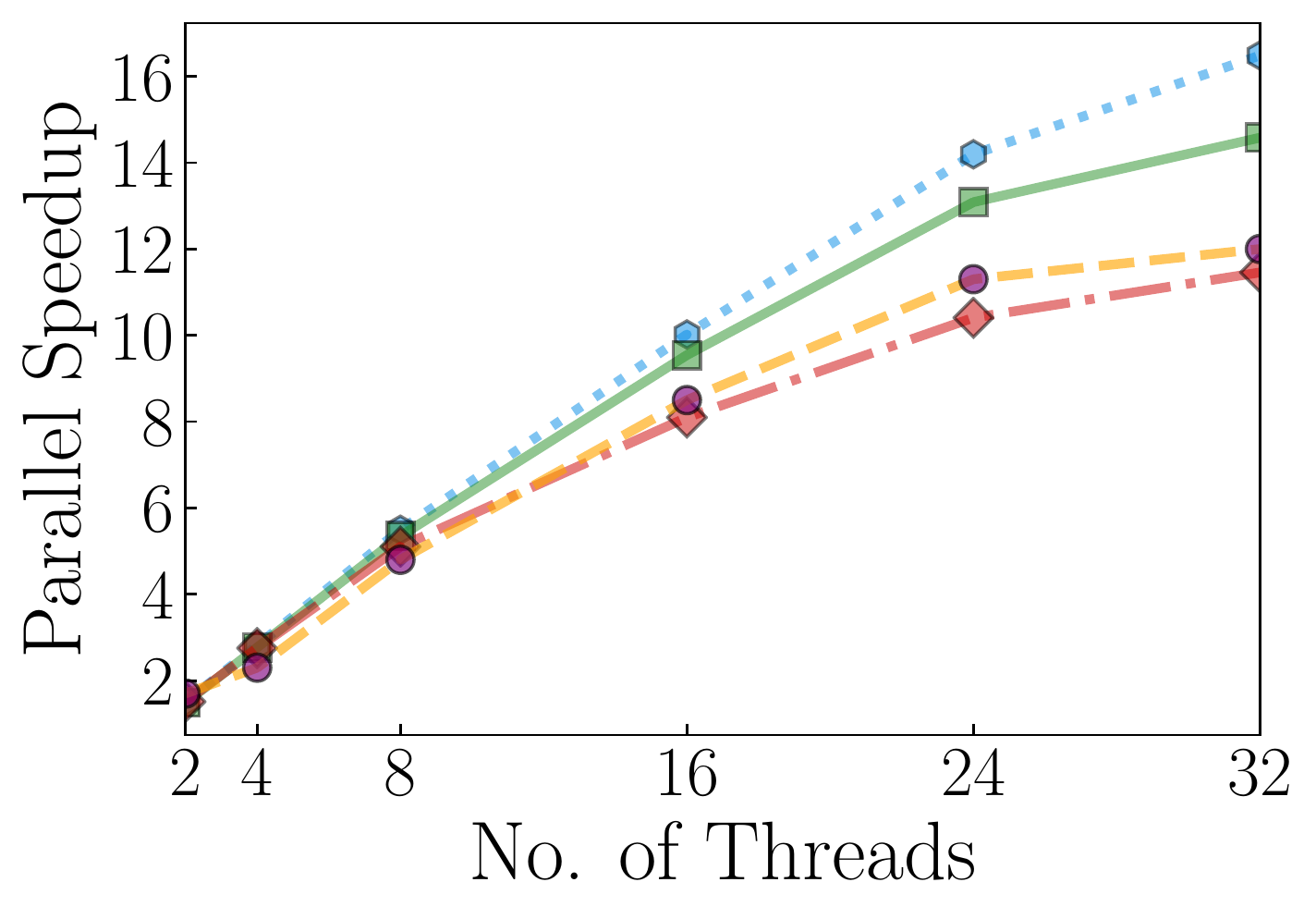}\\
		(a) {\tt dblp-coauthor} &
		(b) {\tt orkut} & 		
		(c) {\tt as-skitter}&
		(d) {\tt wiki-talk}\\
	\end{tabular}
	\caption{\textbf{Parallel speedup of $\parmcedegree$, $\parmcedegen$, $\parmcetriangle$, and $\partomita$ with respect to $\tomita$ as a function of the number of threads. We use total computation time (time for computing ranking + parallel enumeration time) for measuring the speedup.}}
	\label{fig:parallel-speedup-linear-scale}
\end{figure*}

The total runtimes of the parallel algorithms with $32$ threads are shown in Table~\ref{result:parallel-runtimes}. We observe that $\partomita$ achieves a speedup of \textbf{12x} to \textbf{14x} over the sequential algorithm $\tomita$. The three versions of $\parmce$, $\parmcedegree$, $\parmcedegen$, $\parmcetriangle$ achieve a speedup of \textbf{15x} to \textbf{31x} with 32 threads, when we consider only the runtime for maximal clique enumeration. This speedup are smaller for $\parmcedegen$ and $\parmcetriangle$ when we add up the time taken by ranking strategies (See Figure~\ref{fig:parallel-speedup-linear-scale}). 

The reason for the higher runtimes of $\partomita$ when compared with $\parmce$ is the greater cumulative overhead of computing the pivot and in processing the $\cand$ and $\fini$ sets in $\partomita$. For example, for {\tt dblp-coauthors} graph, in $\partomita$, the cumulative overhead of computing $\pivot$ is 248 sec. and cumulative overhead of updating the $\cand$ and $\fini$ is 38 sec. whereas in $\parmce$, these number are 156 sec. and 21 sec. respectively and these reduced cumulative times in $\parmce$ are reflected in the overall reduction in the parallel enumeration time of $\parmce$ over $\partomita$ by a factor of $2$.

\begin{figure*}[htp!]
	\begin{tabular}{c}
		\centering
		\includegraphics[width=0.75\textwidth]{legend-time}
		\vspace{-1ex}
	\end{tabular}
	\centering
	\begin{tabular}{cccc}
		\includegraphics[width=.23\textwidth]{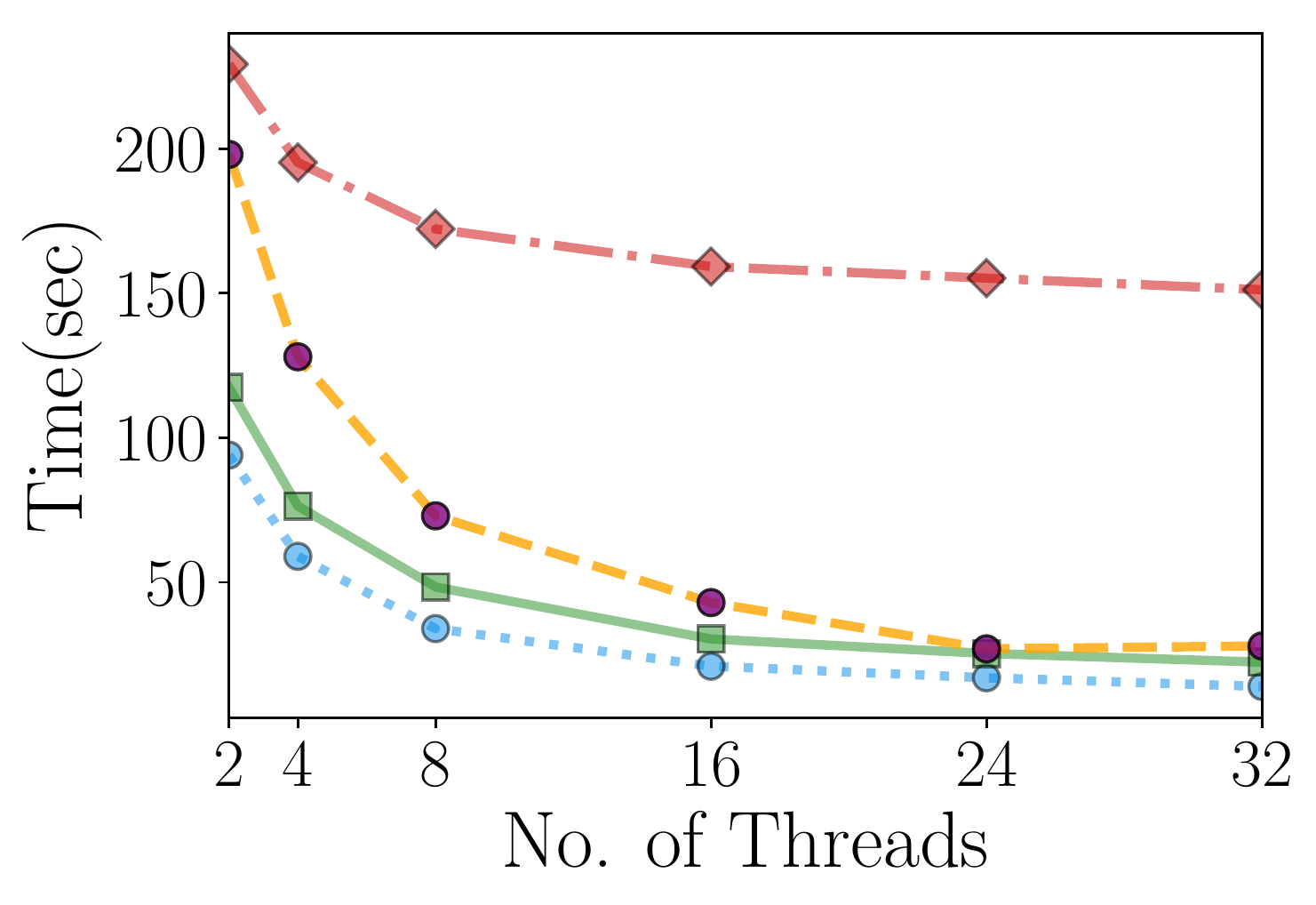} &
		\includegraphics[width=.23\textwidth]{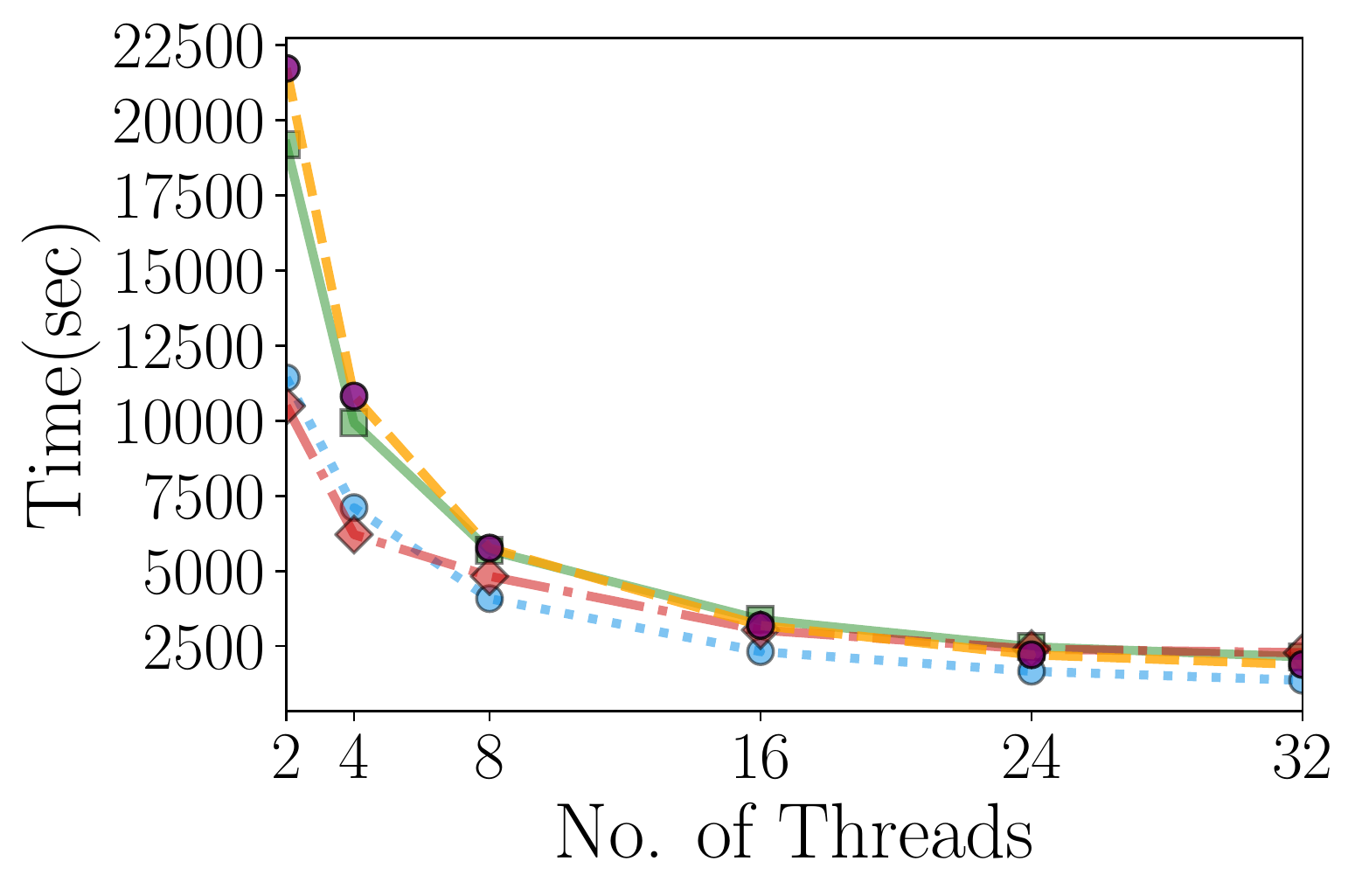} &
		\includegraphics[width=.23\textwidth]{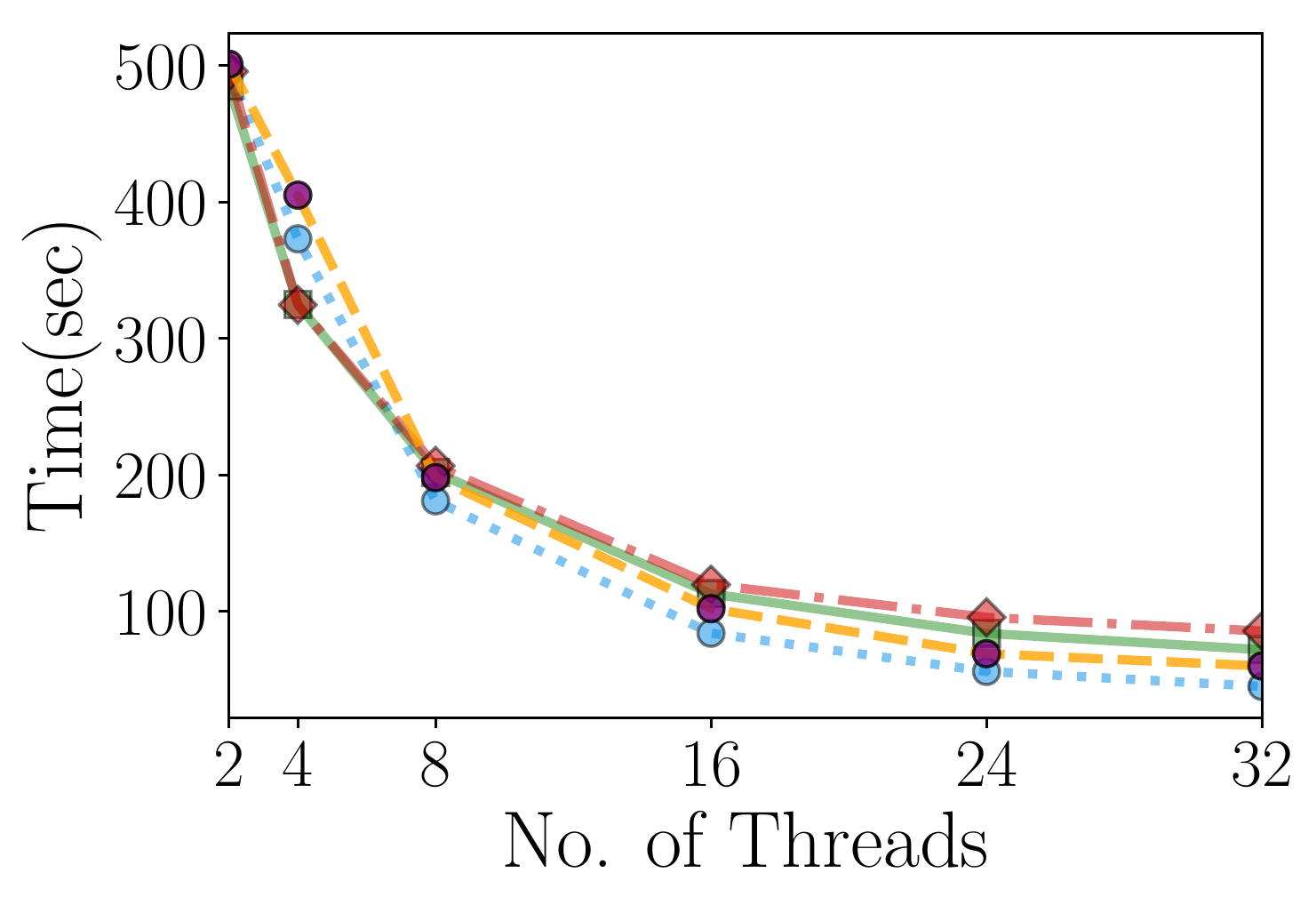} &
		\includegraphics[width=.23\textwidth]{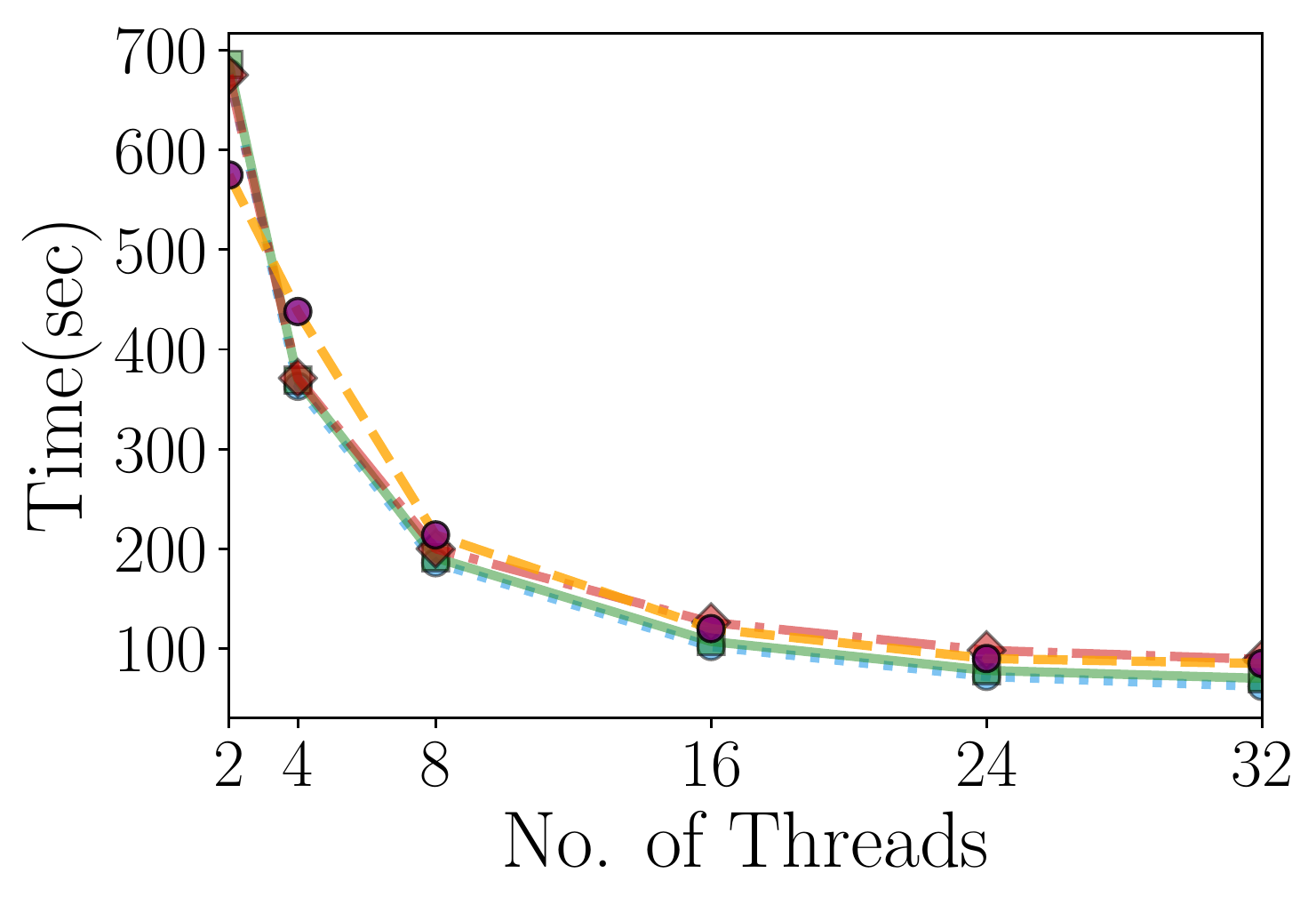}\\
		(a) {\tt dblp-coauthor} &
		(b) {\tt orkut} & 		
		(c) {\tt as-skitter}&
		(d) {\tt wiki-talk}\\
	\end{tabular}
	\caption{\textbf{Total computation time of $\parmcedegree$, $\parmcedegen$, $\parmcetriangle$, and $\partomita$ as a function of the number of threads.}}
	\label{fig:runtimes-per-thread}
\end{figure*}

\begin{figure*}[htp!]
	\centering
	\begin{tabular}{cccc}
		\includegraphics[width=.23\textwidth]{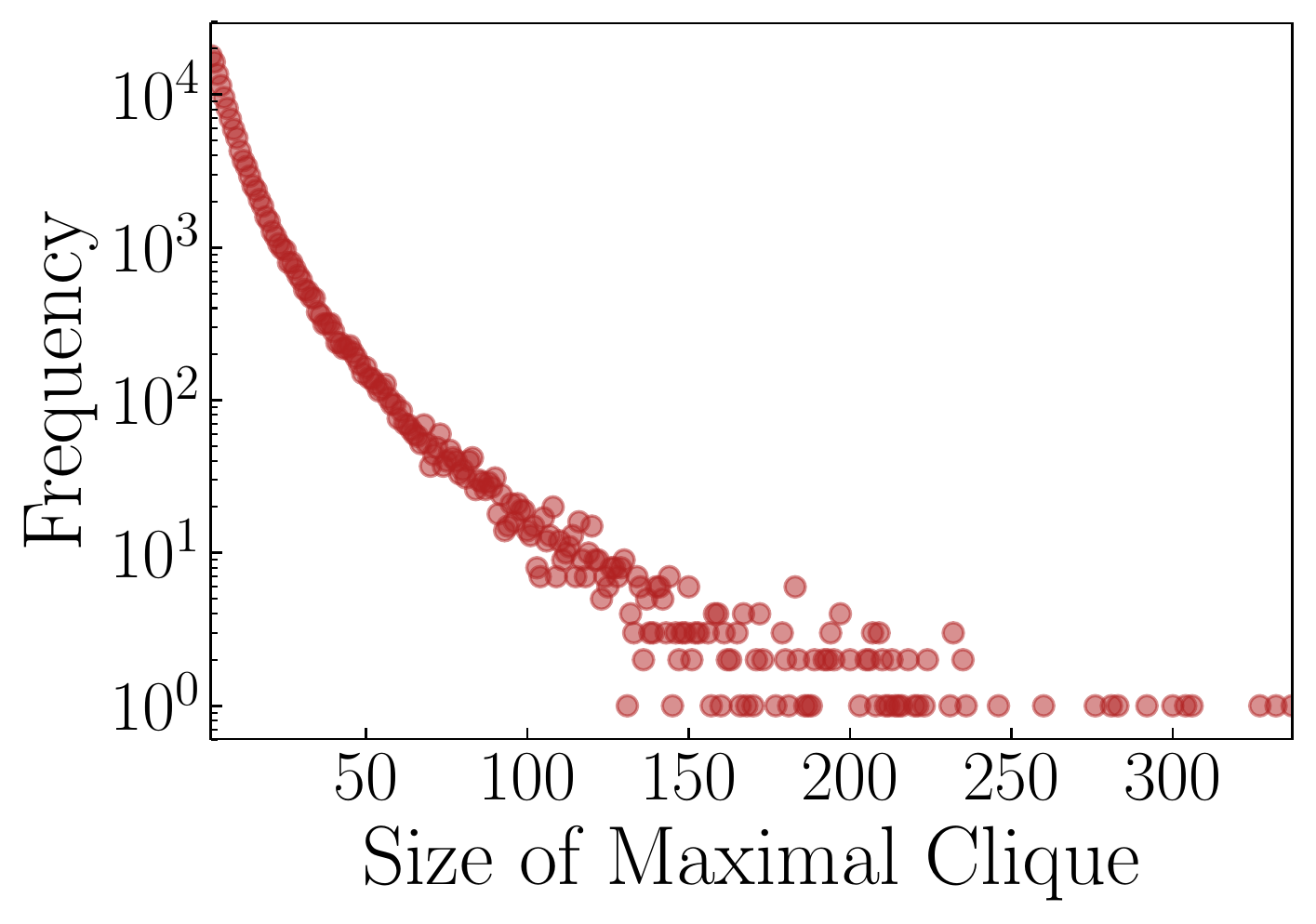} &
		\includegraphics[width=.23\textwidth]{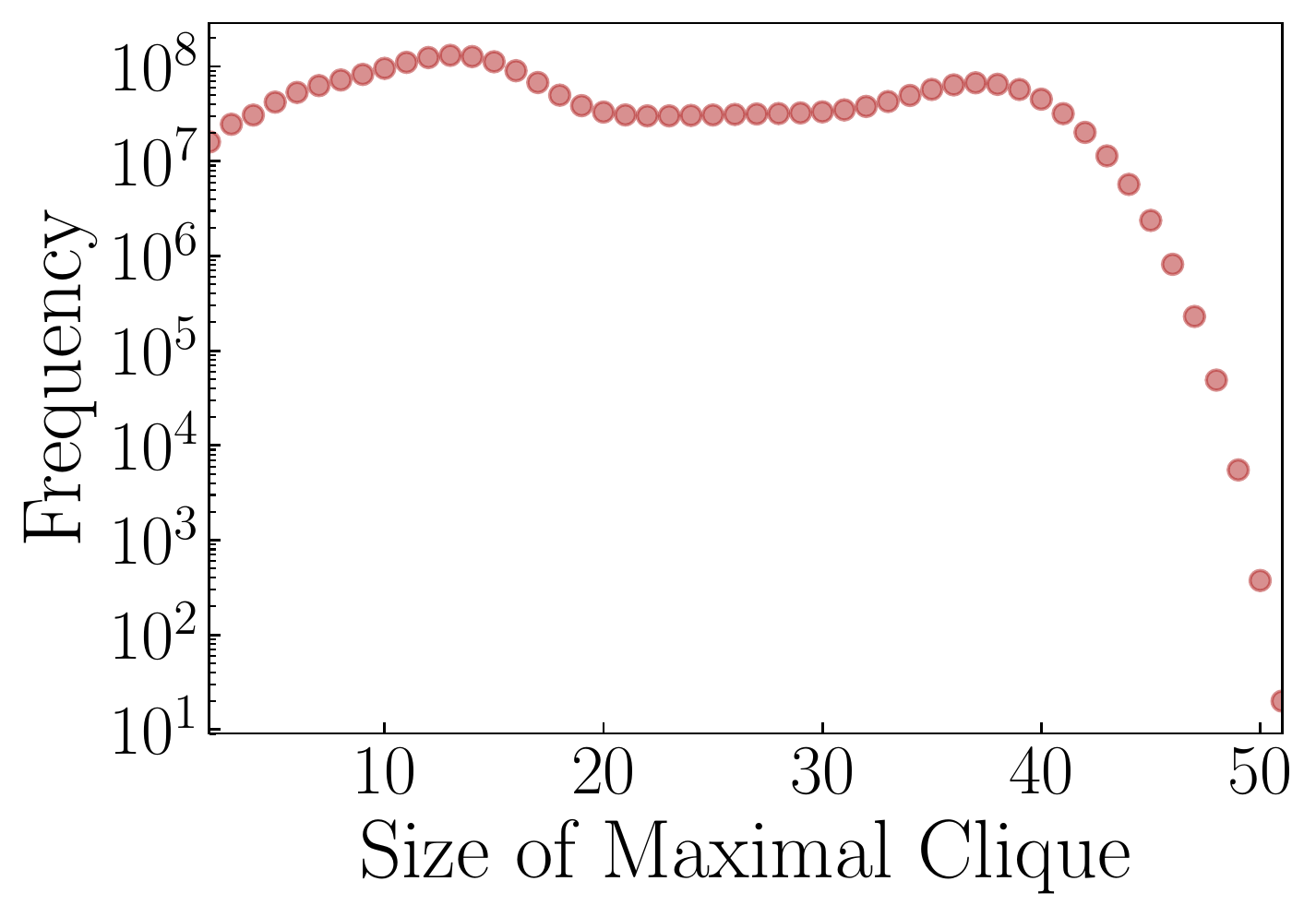} &
		\includegraphics[width=.23\textwidth]{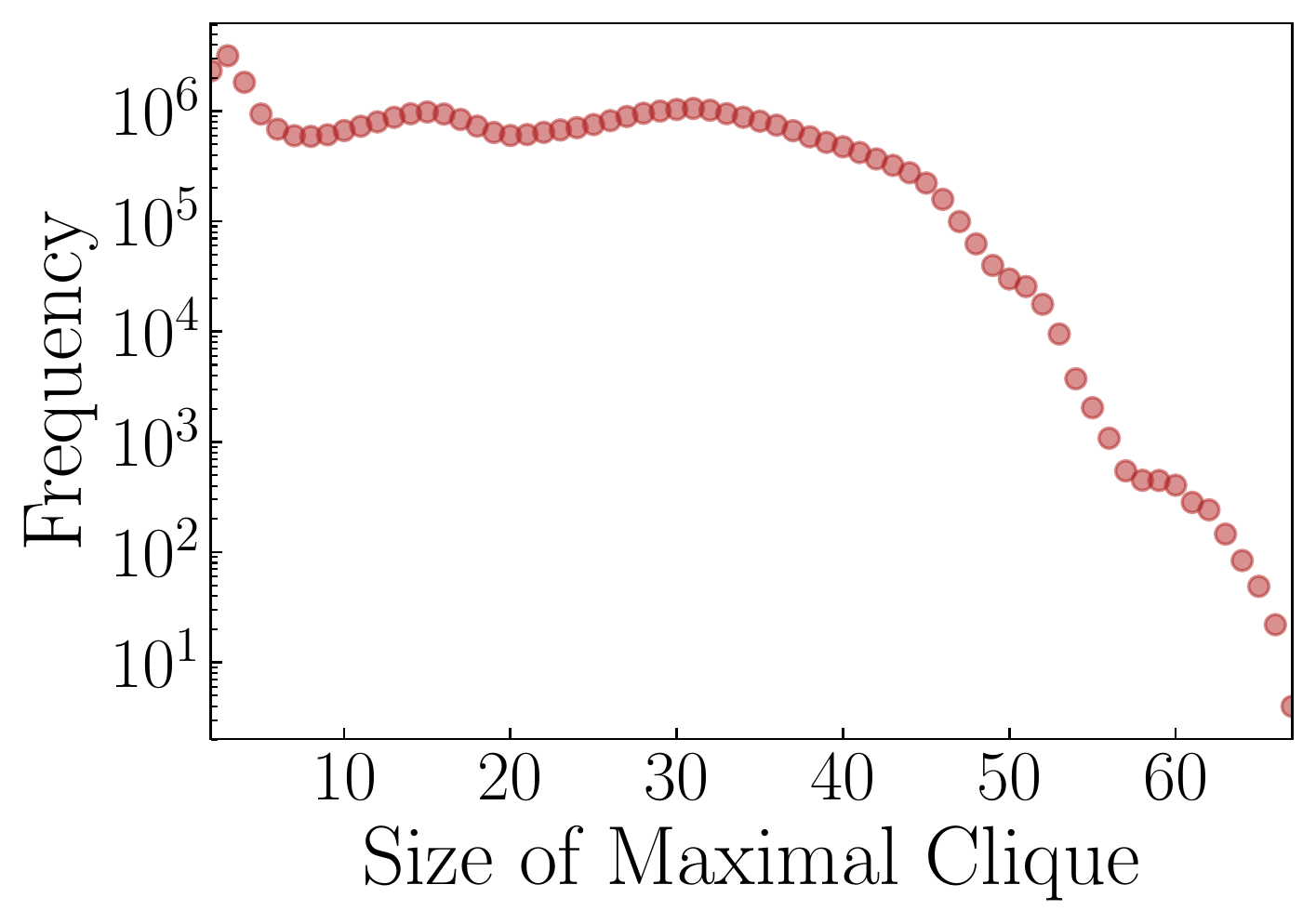} &
		\includegraphics[width=.23\textwidth]{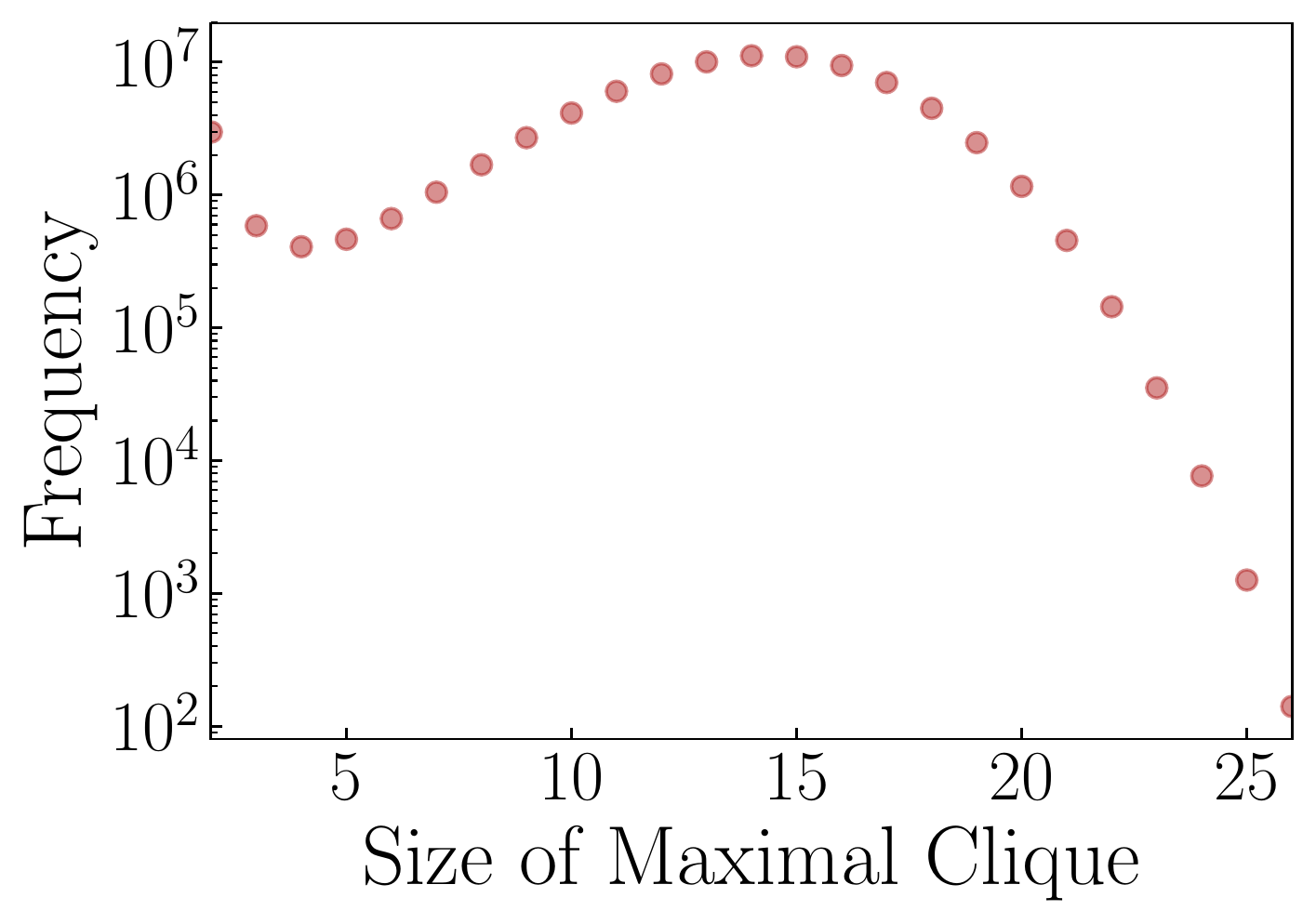}\\
		(a) {\tt dblp-coauthor} &
		(b) {\tt orkut} & 		
		(c) {\tt as-skitter}&
		(d) {\tt wiki-talk}\\
	\end{tabular}
	\caption{\textbf{Frequency distribution of sizes of maximal cliques across different input graphs.}}
	\label{fig:szMC-distribution}
\end{figure*}

\paragraph{Impact of vertex ordering on overall performance of $\parmce$}
Next we consider the influence of different vertex ordering strategies, degree, degeneracy, and triangle count, on the performance of $\parmce$. The total computation time when using different vertex ordering strategies are presented in Table~\ref{result:runtime-splitup}. Overall, we observe that degree based ordering ($\parmcedegree$) usually achieves the smallest (or close to the smallest) runtime for clique enumeration, even when we don't take into account the time to compute the ordering. If we add in the time for computing the ordering, {\em degree based ordering is clearly better than triangle count or degeneracy based orderings}, since degree based ordering is available for free, while the degeneracy based ordering and triangle based ordering require additional computational overhead.

\paragraph{Scaling up with the degree of parallelism}
As the number of threads (and the degree of parallelism) increases, the runtime of $\parmce$ and of $\partomita$ decreases, and the speedup as a function of the number of threads is shown in Figure~\ref{fig:parallel-speedup-linear-scale} and the runtimes are shown in Figure~\ref{fig:runtimes-per-thread}. We see that $\parmcedegree$ achieves a speedup of more than 15x on all graphs, using 32 threads. On the {\tt dblp-coauthors} graph, the speedup with 32 threads was nearly 30x. 

To get a better understanding of the variation of speedups achieved on different input graphs, we plotted the distribution of the sizes of maximal cliques for different input graphs, see Figure~\ref{fig:szMC-distribution}. We observe that the speedup of $\parmce$ is higher on those graphs that have large maximal cliques. For instance, there are many maximal cliques of size in the range $100$ to $330$ for {\tt dblp-coauthors}, and we observed the highest speedup, of nearly 30x with 32 threads, for {\tt dblp-coauthors}. A good speedup of nearly 20x was also observed for the {\tt orkut} graph, which has a large number of maximal cliques, which are of relatively large sizes (the average size of a maximal clique is 20). Overall, we see that the speedup obtained is roughly correlated with the complexity of the graph, measured in terms of the presence of large maximal cliques, as well as the number of such large maximal cliques.

\begin{table*}[htp!]
\centering
\scalebox{0.9}{
\begin{tabular}{| l | r | r | r | r | r | r | r |}
\toprule
\multirow{2}{*}\textbf{DataSet} & \multirow{2}{*}{\parmcedegree} & \multicolumn{3}{|c|}{\parmcedegen} & \multicolumn{3}{|c|}{\parmcetriangle} \\
\cline{3-8}
& & RT & ET & TT & RT & ET & TT \\
\midrule
{\tt dblp-coauthors} & \textbf{14} & 8.4 & 13 & \textbf{21.4} & 138.2 & 14 & \textbf{152.2}\\
\hline
{\tt orkut} & \textbf{1362} & 599.1 & 1542 & \textbf{2141.1} & 786.7 & 1492 & \textbf{2278}\\
\hline
{\tt as-skitter} & \textbf{45} & 26.9 & 45  & \textbf{71.9} & 43.6 & 42 & \textbf{85.6}\\
\hline
{\tt wiki-talk} & \textbf{62} & 8.1 & 62 & \textbf{70.1} & 30.2 & 59 & \textbf{89.2}\\
\bottomrule
\end{tabular}
}
\caption{\textbf{Total computation time (in sec.) of \parmcedegree, \parmcedegen, and \parmcetriangle. ``RT" stands for time for computing the vertex ranking, ``ET" stands for parallel enumeration time, and ``TT" stands for total computation time.}}
\label{result:runtime-splitup}
\end{table*}

\subsection{Comparison with prior work \label{sec:exp-prior}}
We compare the performance of $\parmce$ with prior sequential and parallel algorithms for MCE. We consider the following sequential algorithms: $\greedybb$ due to Segundo et al.~\cite{SAS18}, $\tomita$ due to Tomita et al.~\cite{TTT06}, and $\bkd$ due to Eppstein et al.~\cite{ELS10}. For the comparison with parallel algorithm, we consider algorithm $\cliqueenum$ due to Zhang et al.~\cite{ZA+05}, $\peamc$ due to Du et al.~\cite{DW+09}, \peco due to Svendsen et al.~\cite{SMT+15}, and most recent parallel algorithm $\hashing$ due to Lessley et al.~\cite{LP+17}. The parallel algorithms $\cliqueenum$, $\peamc$, and $\hashing$ are designed for the shared memory model, while $\peco$ is designed for distributed memory. We modified $\peco$ to work with shared memory, by reusing the method for subproblem construction,  and eliminating the need to communicate subgraphs by storing a single copy of the graph in shared memory. We considered three different ordering strategies for $\peco$, which we call $\pecodegree, \pecodegen$, and $\pecotri$. The comparison of performance of \parmce with \peco is presented in Table~\ref{result:compare-with-peco}. We note that $\parmce$ is significantly better than that of $\peco$, no matter which ordering strategy was considered.



\begin{table*}[htp!]
\centering
\begin{tabular}{| l | r | r || r | r || r | r |}
\toprule
\textbf{DataSet} & \pecodegree & \parmcedegree & \pecodegen & \parmcedegen & \pecotri &\parmcetriangle \\
\midrule
{\tt dblp-coauthors} & 73 & \textbf{14} & 78 & \textbf{14} & 74 & \textbf{13}\\
\hline
{\tt orkut} & \num{2001} & \textbf{\num{1362}} & \num{7502} & \textbf{\num{1542}} & \num{2500} & \textbf{\num{1492}} \\
\hline
{\tt as-skitter}  & 272 & \textbf{45}  & 450  & \textbf{45} & 267 & \textbf{42} \\
\hline
{\tt wiki-talk} & \num{1423} & \textbf{62} & \num{1776}  & \textbf{62} & \num{1534} & \textbf{59}  \\
\bottomrule
\end{tabular}
\caption{\textbf{Comparison of parallel enumeration time (in sec.) of $\parmce$ with $\peco$ (modified to use shared memory), using 32 threads. Three different variants are considered for each algorithm, based on the ordering strategy used.}}
\label{result:compare-with-peco}
\end{table*}
\begin{table*}[htp!]
\centering
\scalebox{0.9}{
\begin{tabular}{| r | r | r | r | r |}
\toprule
\textbf{DataSet} & \parmcedegree & \hashing & \cliqueenum & \peamc\\
\midrule
{\tt dblp-coauthors} & 14 & run out of memory in 3 min. & run out of memory in 10 min. & did not complete in 5 hours.\\
\hline
{\tt orkut} & 1362 & run out of memory in 7 min. & run out of memory in 20 min.  & did not complete in 5 hours.\\
\hline
{\tt as-skitter} & 45 & run out of memory in 5 min. & run out of memory in 10 min. & did not complete in 5 hours.\\
\hline
{\tt wiki-talk} & 62 & run out of memory in 10 min. & run out of memory in 20 min. & did not complete in 5 hours.\\
\bottomrule
\end{tabular}
}
\caption{\textbf{Comparison of total computation time (in sec.) of $\parmce$ with $\hashing$.}}
\label{result:compare-with-hashing}
\end{table*}

The comparison of $\parmce$ with other shared memory algorithms $\peamc$, $\cliqueenum$, and $\hashing$ is shown in Table~\ref{result:compare-with-hashing}. The performance of $\parmce$ is seen to be much better than that of any of these prior shared memory parallel algorithms. For the graph {\tt dblp-coauthor}, $\peamc$ did not finish within $5$ hours, whereas $\parmce$ takes around $50$ secs for enumerating  $139$K maximal cliques. The poor running time of  $\peamc$ is due to two following reasons: (1)~the algorithm does not apply efficient pruning techniques such as pivoting, used in $\tomita$, and (2)~the method to determine the maximality of a clique in the search space is not efficient. The $\cliqueenum$ algorithms run out of memory after a few minutes. The reason is that $\cliqueenum$ maintains a bit vector for each vertex that is as large as the size of the input graph, and additionally, needs to store intermediate non-maximal cliques. For each such non-maximal clique, it is required to maintain a bit vector of length equal to the size of the vertex set of the original graph. Therefore, in $\cliqueenum$ a memory issue is inevitable for a graph with millions of vertices.

A recent parallel algorithm in the literature, $\hashing$ also has a significant memory overhead, and ran out of memory on the input graphs that we considered. The reason for its high memory requirement is that $\hashing$ enumerates intermediate non-maximal cliques before finally outputting maximal cliques. The number of such intermediate non-maximal cliques may be very large, even for graphs with few number of maximal cliques. For example, a maximal clique of size $c$ contains $2^c-1$ non-maximal cliques.

\begin{table*}[htp!]
\centering
\scalebox{0.9}{
\begin{tabular}{| l | r | r | r | r | r |}
\toprule
\textbf{DataSet} & \bkd & \greedybb & \parmcedegree & \parmcedegen & \parmcetriangle \\
\midrule
{\tt dblp-coauthors} & 231 & did not finish in 30 min. & 14 & 21.4 & 152.2\\
\hline
{\tt orkut} & \num{19958} & run out of memory in 5 min. & 1362 & 2141.1 & 2278\\
\hline
{\tt as-skitter} &  588 & out of memory in 10 min. & 45 & 71.9 & 85.6\\
\hline
{\tt wiki-talk} & 844 & run out of memory in 10 min. & 62 & 70.1 & 89.2\\
\bottomrule
\end{tabular}
}
\caption{\textbf{Total computation time (sec.) of $\parmce$ (with $32$ threads) and sequential $\bkd$ and $\greedybb$.}}
\label{results:compare-with-sequential}
\end{table*}

Next, we compare the performance of $\parmce$ with that of sequential algorithms $\bkd$ and a recent sequential algorithm $\greedybb$ -- results are in Table~\ref{results:compare-with-sequential}. For large graphs, the performance of $\bkd$ is almost similar to $\tomita$ whereas  $\greedybb$ performs much worse than $\tomita$. Since our $\parmce$ algorithm outperforms $\tomita$, we can conclude that $\parmce$ is significantly faster than other sequential algorithms.


\subsection{Summary of Experimental Results}
We found that both $\partomita$ and $\parmce$ yield significant speedups over the sequential algorithm $\tomita$, sometimes as much as the number of cores available. $\parmce$ using the degree-based vertex ranking always performs better than $\partomita$. The runtime of $\parmce$ using degeneracy/triangle count based vertex ranking is sometimes worse than $\partomita$ due to the overhead of sequential computation of vertex ranking -- note that this overhead is not needed in $\partomita$. The parallel speedup of $\parmce$ is better when the input graph has many large sized maximal cliques. Overall, $\parmce$ consistently outperforms prior sequential and parallel algorithms for MCE.

\remove{
\begin{figure*}[]
	\begin{tabular}{c}
		\centering
		\includegraphics[width=1\textwidth]{plots/vahid/legend-dst}
		\vspace{0ex}
	\end{tabular}
	\centering
	\begin{tabular}{cccc}
		\includegraphics[width=.23\textwidth]{plots/vahid/gcloud/dblp_coauthor-dst-gcloud} &
		\includegraphics[width=.23\textwidth]{plots/vahid/gcloud/orkut-dst-gcloud} &
		\includegraphics[width=.23\textwidth]{plots/vahid/gcloud/as_skitter-dst-gcloud} &
		\includegraphics[width=.23\textwidth]{plots/vahid/gcloud/wiki_talk-dst-gcloud}\\
		(a) {\tt dblp-coauthor} &
		(b) {\tt orkut} & 		
		(c) {\tt as-skitter}&
		(d) {\tt wiki-talk}\\
	\end{tabular}
	\caption{\textbf{Frequency distribution of Parallel Speedup of \partomita for each subproblem (with degree based vertex ordering) over \tomita in different graphs.}}
	\label{fig:speedup-distribution}
\end{figure*}
}

\section{Conclusion}
\label{sec:conclude}
We presented shared memory parallel algorithms for enumerating maximal cliques from a graph. $\partomita$ is a work-efficient parallelization of a sequential algorithm due to Tomita et al.~\cite{TTT06},  and $\parmce$ is a practical adaptation of $\partomita$ that has more opportunities for parallelization and better load balancing. Our algorithms are significant improvements compared with the current state-of-the-art on MCE. Our experiments show that $\parmce$ has a speedup of up to 31x (on a 32 core machine) when compared with an efficient sequential baseline. In contrast, prior shared memory parallel methods for MCE were either unable to process the same graphs in a reasonable time, or ran out of memory. Many questions remain open : (1)~Can these methods scale to even larger graphs, and to machines with larger numbers of cores (2)~How can one adapt these methods to other parallel systems such as a cluster of computers with a combination of shared and distributed memory, or GPUs?


\bibliographystyle{IEEEtran}
\bibliography{IEEEabrv,cliques}

\end{document}